\documentclass[12pt]{article}
\usepackage{epsf, cite, amssymb}
\usepackage{epsfig}
\makeatletter
\renewcommand\section{\@startsection {section}{1}{\z@}%
                                  {-3.5ex \@plus -1ex \@minus -.2ex}
                                  {2.3ex \@plus.2ex}%
                                  {\normalfont\large\bfseries}}
\renewcommand\subsection{\@startsection{subsection}{2}{\z@}%
                                    {-3.25ex\@plus -1ex \@minus -.2ex}%
                                    {1.5ex \@plus .2ex}%
                                    {\normalfont\bfseries}}
\makeatother

\def\IZ{\relax\ifmmode\mathchoice
{\hbox{\cmss Z\kern-.4em Z}}{\hbox{\cmss Z\kern-.4em Z}}
{\lower.9pt\hbox{\cmsss Z\kern-.4em Z}} {\lower1.2pt\hbox{\cmsss
Z\kern-.4em Z}}\else{\cmss Z\kern-.4em Z}\fi}
\def\IR{\relax{\rm I\kern-.18em R}}

\def\one{{\hbox{ 1\kern-.8mm l}}}

\newlength{\bredde}
\def\slash#1{\settowidth{\bredde}{$#1$}\ifmmode\,\raisebox{.15ex}{/}
\hspace*{-\bredde} #1\else$\,\raisebox{.15ex}{/}\hspace*{-\bredde}
#1$\fi}

\newcommand {\Cbar}
   {\mathord{\setlength{\unitlength}{1em}
    \begin{picture}(0.6,0.7)(-0.1,0)
       \put(-0.1,0){\rm C}
       \thicklines
       \put(0.2,0.05){\line(0,1){0.55}}
    \end {picture}}}
\newsavebox{\zzzbar}
\sbox{\zzzbar}
 {\setlength{\unitlength}{0.9em}
 \begin{picture}(0.6,0.7)
 \thinlines
 \put(0,0){\line(1,0){0.6}}
 \put(0,0.75){\line(1,0){0.575}}
 \multiput(0,0)(0.0125,0.025){30}{\rule{0.3pt}{0.3pt}}
 \multiput(0.2,0)(0.0125,0.025){30}{\rule{0.3pt}{0.3pt}}
 \put(0,0.75){\line(0,-1){0.15}}
 \put(0.015,0.75){\line(0,-1){0.1}}
 \put(0.03,0.75){\line(0,-1){0.075}}
 \put(0.045,0.75){\line(0,-1){0.05}}
 \put(0.05,0.75){\line(0,-1){0.025}}
 \put(0.6,0){\line(0,1){0.15}}
 \put(0.585,0){\line(0,1){0.1}}
 \put(0.57,0){\line(0,1){0.075}}
 \put(0.555,0){\line(0,1){0.05}}
 \put(0.55,0){\line(0,1){0.025}}
 \end{picture}}
\newcommand{\Zbar}{\mathord{\!{\usebox{\zzzbar}}}}

\newcommand{\ena}{\end{eqnarray}}
\newcommand{\beqa}{\begin{eqnarray}}
\newcommand{\eeqa}{\end{eqnarray}}
\newcommand{\bea}{\begin{eqnarray}}
\newcommand{\eea}{\end{eqnarray}}

\newcommand{\eq}[1]{(\ref{#1})}

\newcommand{\be}{\begin{equation}}
\newcommand{\ee}{\end{equation}}

\newcommand{\half}{{1\over2}}
\newcommand{\Tr}{{\rm Tr}}
\usepackage{graphicx}

\def\be{\begin{equation}}
\def\ee{\end{equation}}
\def\beq{\begin{eqnarray}}
\def\eeq{\end{eqnarray}}

\def\({\left (}
\def\){\right )}
\def\[{\left [}
\def\[{\right ]}

\def\ba{\begin{eqnarray}}
\def\ea{\end{eqnarray}}

\input amssym.def
\input amssym.tex
\begin{document}
\begin{titlepage}
\begin{flushright}
arXiv:0905.0709
\end{flushright}
\vfill
\begin{center}
{\LARGE\bf A multitrace deformation of ABJM theory} \\
\vskip 10mm

{\large Ben Craps,$^{1,2}$ Thomas Hertog$^{2,3}$ and Neil Turok$^{4}$}
\vskip 7mm

{\em $^1$ Theoretische Natuurkunde, Vrije Universiteit Brussel, \\
Pleinlaan 2, B-1050 Brussels, Belgium}
\vskip 3mm 
{\em 
$^2$International Solvay Institutes, Boulevard du Triomphe, \\
ULB--C.P.231, B-1050 Brussels, Belgium}
\vskip 3mm 
{\em 
$^3$ APC, Universit\'e Paris 7, 10 rue A.Domon et L.Duquet, 75205 Paris, France\\}
\vskip 3mm 
{\em 
$^4$ Perimeter Institute for Theoretical Physics,
31 Caroline St N,
Waterloo,\\ Ontario N2L2Y5,
Canada, and\\ 
DAMTP, CMS, Wilberforce Road, Cambridge, CB3 0WA, UK} 

\vskip 3mm
{\small\noindent  {\tt Ben.Craps@vub.ac.be, Thomas.Hertog@apc.univ-paris7.fr, nturok@perimeterinstitute.ca}}
\end{center}
\vfill

\begin{center}
{\bf ABSTRACT}\vspace{3mm}
\end{center}

Motivated by the study of big crunch singularities in asymptotically $AdS_4$ spacetimes, we consider a marginal triple trace deformation of ABJM theory. The deformation corresponds to adding a potential which is unbounded below. In a 't~Hooft large $N$ limit, the beta function for the triple trace deformation vanishes, which is consistent with the near-boundary behavior of the bulk fields. At the next order in the $1/N$ expansion, the triple trace couplings exhibit non-trivial running, which we analyze explicitly in the limit of zero 't~Hooft coupling, in which the model reduces to an  $O(N)\times O(N)$ vector model with large $N$. In this limit, we establish the existence of a perturbative UV fixed point, and we comment on possible non-perturbative effects. We also show that the bulk analysis leading to big crunch singularities extends to the $\Zbar_k$ orbifold models dual to ABJM theory.

\vfill

\end{titlepage}
\section{Introduction and summary}

M-theory compactified on $S^7$ with asymptotically $AdS_4$ boundary conditions allows a consistent truncation to four-dimensional supergravity with a negative cosmological constant and a single scalar field whose negative mass squared lies just above the Breitenlohner-Freedman bound. Besides the usual supersymmetric boundary conditions, there is a different set of possible, well-defined boundary conditions that break supersymmetry but preserve all AdS-symmetries. In \cite{Hertog:2004rz} it was shown that the theory with non-supersymmetric, AdS-invariant boundary conditions admits solutions where smooth, asymptotically AdS initial data evolve into a big crunch singularity -- a spacelike singularity that reaches the boundary of $AdS_4$ in finite global time. 

The holographic dual to M-theory in asymptotically $AdS_4\times S^7$ spacetimes is the three-dimensional superconformal field theory that describes the low energy dynamics of coincident M2-branes~\cite{Maldacena:1997re}. This theory can be thought of as living on the boundary of $AdS_4$. Adopting non-supersymmetric but AdS-invariant boundary conditions for the bulk corresponds to adding a marginal triple trace potential to the boundary theory. With this correction, the tree level potential of the boundary theory no longer has a minimum, indicating that the Hamiltonian of the quantum boundary theory may be unbounded below. In Refs. \cite{Hertog:2004rz, Hertog:2005hu}, the suggestion was made that one might be able to learn something about cosmological singularities in the bulk by studying field theories with potentials which are unbounded below.   

At the time of \cite{Hertog:2004rz, Hertog:2005hu}, however, not much was known about the M2-brane theory, even without the unstable deformation. It arises as the infrared (strong coupling) limit of the super-Yang-Mills (SYM) theory living on D2-branes, but this infrared limit was hard to describe explicitly. For instance, its spectrum of chiral operators was derived not through field theory computations, but by using the AdS/CFT correspondence and the known Kaluza-Klein spectra of eleven-dimensional supergravity compactified on $S^7$ \cite{Aharony:1998rm}. But without explicit knowledge of the dual theory, one could not perform reliable field theory computations to give information about cosmological singularities.

A more specific criticism of \cite{Hertog:2004rz} was raised in \cite{Elitzur:2005kz}, where it was argued, based on an analogy with the $O(N)$ vector model at large $N$, that the deformation of the conformal field theory is marginally irrelevant. Since the behavior of the potential for large field values would then depend on an unknown ultraviolet completion of the theory, it was argued that the unbounded below nature of the potential might be an artifact of the tree-level approximation and in particular could be absent in the full quantum theory.

For these reasons, we have recently studied related $AdS_5\times S^5$ models \cite{Craps:2007ch}, also suggested in \cite{Hertog:2004rz}. In these models, the undeformed dual field theory is ${\cal N}=4$ SYM in four dimensions, which is very well understood. The deformation corresponds to adding a negative, unbounded double trace potential \cite{Aharony:2001pa, Witten:2001ua, Berkooz:2002ug}. In this theory, the coupling of the negative double trace deformation is asymptotically free in the large $N$ limit \cite{Witten:2001ua}, which we used to argue that the quantum effective potential is unbounded below (the argument for a single scalar field was given in \cite{Coleman:1973sx}). The relevant coupling becomes arbitrarily small in the regime of interest for studying the cosmological singularity (namely large fields in the boundary theory), rendering perturbation theory more and more reliable as the singularity approaches.

Recently, a concrete proposal for the theory of $N$ coincident M2-branes was put forward in Ref. \cite{Aharony:2008ug} (ABJM). The ABJM theory is an ${\cal N}=6$ superconformal $U(N)\times U(N)$ Chern-Simons theory with levels $k$ and $-k$, respectively. For $k>1$, the M2-branes are localized at the fixed point of a $\Zbar_k$ orbifold of Minkowski space. The presence of the two parameters $N$ and $k$ allows one to define a 't~Hooft limit $N\to\infty$ with $N/k$ fixed. In this regime, one of the $S^7$ dimensions in the eleven-dimensional bulk becomes small and the theory is best described by type IIA string theory on $AdS_4\times \Cbar P_3$.  

In the present paper, we revisit the issues mentioned above in the context of the ABJM theory. Unlike the model studied in\cite{Craps:2007ch}, in this case the triple trace potential does not have a definite sign, so the $O(N)$ vector model analogy does not apply directly. Therefore, we introduce a tri-critical $O(N)\times O(N)$ vector model as a better analogue and study its fixed point structure -- in fact, in the limit of weak 't~Hooft coupling ($N/k\to 0$), our deformation of ABJM theory precisely reduces to the $O(2 N^2)\times O(2 N^2)$ vector model at large $N$. In this limit, we find that the perturbative beta functions for the various sextic couplings vanish. (A similar scale-independence at strong 't~Hooft coupling can be inferred from the near-boundary behavior of the corresponding bulk scalar \cite{Hertog:2004rz}.) At the next order in the $1/N$ expansion, the beta functions are non-trivial and the $O(N)\times O(N)$ vector model has several non-trivial UV fixed points, one of which corresponds to the UV regime of our potential with indefinite sign. Modulo subtleties we are about to mention, this shows that at least at zero 't~Hooft coupling, as in the negative double trace deformation of ${\cal N}=4$ SYM case, the coupling of a negative triple trace deformation is asymptotically free and the quantum effective potential is unbounded below. Parenthetically, let us mention that large $N$ non-perturbative effects are known to destabilize the model in the UV when a time-independent, static system is considered (see \cite{Bardeen:1983rv} for the $O(N)$ vector model, and \cite{Rabinovici:1987tf} for the $O(N)\times O(N)$ vector model). However, preliminary analysis indicates that these instabilities are in fact consistent with the time-dependent, cosmological applications we have in mind~\cite{CHTinprogress}.

In this paper, we also extend the bulk analysis of \cite{Hertog:2004rz} (which would correspond to $k=1$) to $k>1$, which is necessary to make contact with the 't~Hooft regime in which we do the field theory analysis. In particular, we show that the scalar field present in the consistent truncation in \cite{Hertog:2004rz} survives the $\Zbar_k$ orbifolding, so that the four-dimensional analysis of \cite{Hertog:2004rz} also holds for $k>1$. 

With these results in hand, one can attempt to define unitary evolution in these theories by using self-adjoint extensions. It will then be interesting to study the implications of this for the nature of cosmological singularities in the bulk. The results of this work will appear elsewhere~\cite{CHTinprogress}.

The structure of this paper is as follows. In section~\ref{HH}, we briefly review the work of \cite{Hertog:2004rz} on big crunch solutions of AdS supergravity coupled to a scalar field with non-supersymmetric boundary conditions. In section~\ref{ABJM}, we review the ABJM theory of coincident M2-branes and its gravity dual. In section~\ref{def}, we propose a triple trace deformation dual to the modified boundary conditions in the bulk. By studying the properties of the $O(N)\times O(N)$ vector model under renormalization group flow, we show that, at least in the weak 't~Hooft coupling limit, the couplings of the triple trace potential have a perturbative UV fixed point. We discuss $\Zbar_k$ orbifolds of the bulk models of \cite{Hertog:2004rz} and show that the scalar field of interest survives the orbifolding.   

\setcounter{equation}{0}
\section{AdS cosmology}
\label{HH}
M-theory in asymptotically $AdS_4 \times S^7$ spacetimes has $D=4,\, {\cal N}=8$ gauged supergravity as its low energy limit. This theory contains the graviton, 28 gauge bosons in the adjoint of SO(8) and 70 real scalars as its bosonic degrees of freedom. It allows a consistent truncation to four-dimensional gravity coupled to a single scalar field\cite{Duff:1999gh}:
\be\label{trunc}
S=\int d^4x \sqrt{g}\left[ {1\over 2} R-{1\over 2}(\nabla\varphi)^2+{1\over R_{AdS}^2}\left(1+2 \cosh(\varphi)\right) \right],
\ee
where we have chosen units in which the 4d Planck mass is unity.\footnote{This truncation corresponds to setting $\phi^{(12)}=0$, $\phi^{(13)}=\phi^{(14)}=\varphi$ and the gauge fields to zero in equation (2.11) of \cite{Duff:1999gh}. As can be seen from (2.12) of \cite{Duff:1999gh}, this choice preserves $SO(4)\times SO(2)\times SO(2)$. Truncations preserving $SO(6)\times SO(2)$ and $SO(4)\times SO(4)$ are also possible.} The maximum of the potential at $\varphi=0$ corresponds to the $AdS_4$ vacuum solution. Small fluctuations around the maximum of the potential have $m^2=-2 R_{AdS}^{-2}$, which is above the Breitenlohner-Freedman bound $m_{BF}^2=-{9\over 4} R_{AdS}^{-2}$ \cite{Breitenlohner82}. Hence, the $AdS_4$ solution is perturbatively stable. The positive mass theorem implies that, with the usual supersymmetric boundary conditions, it is also non-perturbatively stable. As we shall now review, this need not be the case with other AdS-invariant boundary conditions~\cite{Hertog:2004rz}.

In global coordinates, the $AdS_4$ metric reads
\be
ds^2=R_{AdS}^2\left( -(1+r)^2dt^2+{dr^2\over 1+r^2}+r^2d\Omega_2 \right).
\ee
In asymptotically AdS solutions, the scalar field $\varphi$ decays at large radial coordinate as
\be \label{asymp}
\varphi\sim{\alpha(t,\Omega)\over r}+{\beta(t,\Omega)\over r^2}.
\ee
The usual boundary conditions correspond to taking either $\alpha=0$ (which can be chosen for any $m^2$) or $\beta=0$ (which can be chosen for scalars in the mass range $-{9\over 4} R_{AdS}^{-2} <m^2< -{5\over 4} R_{AdS}^{-2}$). These are in fact two special cases out of a one-parameter class of boundary conditions that are anti-de Sitter invariant and allow the construction of well-defined and finite Hamiltonian generators for all elements of the anti-de Sitter algebra \cite{Hertog:2004dr,Henneaux2006}. The more general boundary conditions are given by
\be\label{bc}
\beta=-h\alpha^2,
\ee
where $h$ is an arbitrary constant \cite{Hertog:2004dr}. For $h>0$, solutions were found \cite{Hertog:2004rz} in which smooth asymptotically AdS initial data evolved into a big crunch, a spacelike singularity reaching the boundary of AdS in finite global time.%
\footnote{\label{footh}
The restriction to $h>0$ is not essential. In the solutions of \cite{Hertog:2004rz}, the $\alpha$ coefficient of the initial profile of the bulk scalar field is positive. Since the potential of the bulk scalar field in \eq{trunc} is even in $\varphi$, there exist similar solutions where $\alpha$ is negative in the initial profile, and $\beta$ positive. These are solutions with $h<0$, for which the scalar field becomes negative towards the interior of the bulk. 
}

M-theory in asymptotically $AdS_4 \times S^7$ spacetimes with $\beta=0$ boundary conditions is dual to the three-dimensional superconformal field theory describing the low energy dynamics of coincident M2-branes. The scalar mode $\alpha(t,\Omega)$ of a bulk solution corresponds to the expectation value of the dual operator ${\cal O}$ in the boundary theory. (Choosing a fixed $\beta\neq 0$ would correspond to adding a source term $\int\beta {\cal O}$ to the action of the boundary theory.) The bulk scalar field $\varphi$ is dual to an operator ${\cal O}$ of dimension 1, transforming in the traceless symmetric two-tensor representation of SO(8). In general, adding a term $-\int W({\cal O})$ to the action of the dual field theory corresponds in the bulk theory to adopting modified boundary conditions $\beta (\alpha)$ such that $\beta=W'(\alpha)$ \cite{Witten:2001ua,Berkooz:2002ug}. Taking boundary conditions \eq{bc} is therefore dual to adding a marginal triple trace operator to the boundary action:
\be\label{deform}
S\rightarrow S+{h\over 3}\int{\cal O}^3.
\ee
In section~\ref{def}, we shall see that the operator ${\cal O}$ can take arbitrarily large positive and negative values, so that the potential we have added is unbounded below for any non-zero value of $h$. This is consistent with our earlier comment in footnote~\ref{footh}.  

Unlike the asymptotically $AdS_5\times S^5$ model studied in \cite{Craps:2007ch}, the asymptotic behavior \eq{asymp} with \eq{bc} does not involve a logarithmic dependence on the radial coordinate.\footnote{We note that the absence of a $\varphi^3$ interaction in the potential in \eq{trunc} is a necessary condition for there to be no logarithmic tails in the asymptotic profile of $m^2=-2$ scalars \cite{Henneaux2006}.} In the dual field theory, this corresponds to the fact that conformal invariance is preserved to leading order in $1/N$, which we shall discuss in section~\ref{def}.
\setcounter{equation}{0}
\section{ABJM theory}
\label{ABJM}
Recently, Aharony, Bergman, Jafferis and Maldacena (ABJM) introduced an ${\cal N}=6$ superconformal $U(N)\times U(N)$ Chern-Simons-matter theory with levels $k$ and $-k$, respectively. The two $U(N)$ gauge fields are denoted $A_\mu$ and $\hat A_\mu$. The theory contains scalar fields $Y^A,\ A=1,\ldots 4$ transforming in the fundamental representation of an $SU(4)$ R-symmetry. (Here, we are using the notation of \cite{Benna:2008zy}.) Each $Y^A$ transforms in the bifundamental $(N,\bar N)$ representation of the gauge group. The Hermitean conjugate scalar fields $Y_A^\dagger$ transform in the anti-fundamental representation of $SU(4)$ and the $(\bar N, N)$ of the gauge group. We will not need the fermionic fields explicitly in this paper.

The action reads
\ba\label{ABJMaction}
S&=&\int d^3x\left[
{k\over 4\pi}\epsilon^{\mu\nu\lambda}\Tr(A_\mu\partial_\nu A_\lambda+{2i\over3} A_\mu A_\nu A_\lambda-\hat A_\mu\partial_\nu \hat A_\lambda-{2i\over3} \hat A_\mu \hat A_\nu \hat A_\lambda)\right.\nonumber\\
&&\ \ \ \ \ \ \ \ \ \ \left. -\Tr(D_\mu Y^A)^\dagger D^\mu Y^A + V_{\rm bos} + \ {\rm terms\ with\ fermions}
\right],
\ea
with
\ba \label{ABJMpot}
V_{\rm bos}&=& -{4\pi^2\over 3k^2}\Tr \left[Y^AY_A^\dagger Y^B Y_B^\dagger Y^C Y_C^\dagger+ Y_A^\dagger Y^A Y_B^\dagger Y^B Y_C^\dagger Y^C
\right.\nonumber\\
&& \left.
\ \ \ \ \ \ \ \ \ \ \ \ \ +4Y^AY_B^\dagger Y^C Y_A^\dagger Y^B Y_C^\dagger-6Y^AY_B^\dagger Y^B Y_A^\dagger Y^CY_C^\dagger
\right].
\ea

The proposal of \cite{Aharony:2008ug} is that this theory is the world-volume action for $N$ coincident M2-branes on a $\Zbar_k$ orbifold of $\Cbar^4$, with the generator of $\Zbar_k$ acting as 
\be\label{zkaction}
y^A\rightarrow \exp(2\pi i/k)\, y^A
\ee 
on complex coordinates $y^A$. The coupling constant of the ABJM theory is $1/k$. We will be interested in the ``'t~Hooft'' limit of large $N$ with $N/k$ fixed. In this limit, the theory is weakly coupled for $k\gg N$ and strongly coupled for $k\ll N$.

The gravity dual of this system of M2-branes is a $\Zbar_k$ orbifold of $AdS_4\times S^7$. Before orbifolding, the $AdS_4\times S^7$ solution of eleven-dimensional supergravity with $N'$ units of four-form flux reads
\ba
ds^2&=&{R^2\over 4}ds^2_{AdS_4}+R^2ds^2_{S^7};      \label{cp3} \\
F_4&\sim&N'\epsilon_4;\\
{R\over l_p}&=&(32\pi^2N')^{1/6},
\ea
where $ds^2_{AdS_4}$ and $ds^2_{S^7}$ have unit radius.

The $\Zbar_k$ identification, which acts on the $S^7$ as in \eq{zkaction}, preserves an $SU(4) \times U(1)$ subgroup of the isometry group of $S^7$. It is convenient to rewrite the unit seven-sphere as an $S^1$ fibration over $\Cbar P^3$:
\be
ds_{S^7}^2=(d\chi+\omega)^2+ds_{CP^3}^2,
\ee
where $\chi$ has period $2\pi$ and $\omega$ is a connection on a topologically non-trivial $U(1)$ bundle on $\Cbar P^3$ \cite{Nilsson:1984bj}. The $\Zbar_k$ identification simply changes the period of $\chi$ to $2\pi/k$. In order to have $N$ units of flux on the quotient space, we choose $N'=kN$. While the radius of the $\Cbar P^3$ factor in \eq{cp3} is always large in Planck units if $kN\gg1$, the radius of the $\chi$ circle in Planck units is of order $R/kl_p\sim(kN)^{1/6}/k$, which is very small in the 't~Hooft limit. Therefore, the appropriate description in this regime is as a weakly coupled type IIA string theory. The radius of curvature in string units turns out to be of order $(N/k)^{1/4}$, so the bulk is stringy when the 't~Hooft coupling is small.

\setcounter{equation}{0}
\section{A triple trace deformation of ABJM theory}
\label{def}

The consistent truncation \eq{trunc} of $D=4,\, {\cal N}=8$ gauged supergravity was introduced in \cite{Duff:1999gh}. The bulk scalar $\varphi$, which corresponds to a specific quadrupole deformation of $S^7$ and transforms as a symmetric traceless tensor under $SO(8)$, is invariant under independent $U(1)$ rotations of the four complex coordinates $y^A$ (see Eq.~(2.9) of \cite{Duff:1999gh}), and in particular under the identification \eq{zkaction}. This implies that $\varphi$ survives the $\Zbar_k$ quotient, so that the bulk analysis of \cite{Hertog:2004rz} extends to $k>1$, in particular to the 't~Hooft limit of interest in the present paper.
\begin{figure}
\begin{center}
\epsfig{file=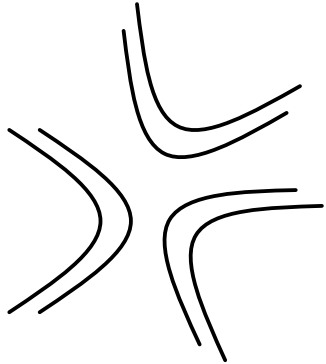, width=6cm}
\end{center}
\caption{The $\left[{\rm Tr}(YY^\dagger)\right]^3$ vertex in double line notation.}
\label{vertex6}
\end{figure}
The operator ${\cal O}$ is a dimension one chiral primary operator with the same symmetry properties as $\varphi$ under the preserved $SU(4)$ subgroup of $SO(8)$. A natural candidate is
\be\label{calO}
{\cal O}={1\over N^2}\Tr(Y^1Y_1^\dagger-Y^2Y_2^\dagger).
\ee
To understand the factor $1/N^2$ in \eq{calO}, note that in general the large $N$ limit of theories of matrix-valued fields $\Phi$ is taken as follows (see for instance \cite{Witten:2001ua}). Trace operators are normalized as ${\cal O}=\Tr F(\Phi)/N$ and the action has the form $N^2 W({\cal O})$, where neither $F$ nor $W$ depend explicitly on $N$. The fields $Y$ appearing in the action \eq{ABJMaction} are rescaled to have an $N$-independent kinetic term in the 't~Hooft limit: $Y\sim \sqrt N \Phi$, which explains the extra factor of $1/N$ in \eq{calO}. The triple trace vertex appearing in the deformation \eq{deform} is drawn in 't~Hooft double line notation in figure~\ref{vertex6}. 
In terms of a coupling $f$ that is kept fixed as the 't~Hooft limit is taken, we have added to the single trace potential \eq{ABJMpot} a triple trace term
\be \label{triple}
V=-{f\over N^4}\left[\Tr(Y^1Y_1^\dagger-Y^2Y_2^\dagger)\right]^3,
\ee
where the $1/N^4$ arises from the $N^2$ in front of the action and a $1/N^6$ from \eq{calO}.

\begin{figure}
\begin{center}
\epsfig{file=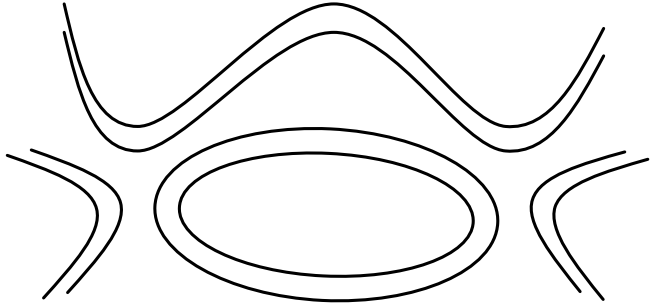, width=10cm}
\end{center}
\caption{Two-loop diagram that renormalizes the coupling $f$ at order $1/N^2$.}
\label{2loop}
\end{figure}

Note that the potential \eq{triple} is unbounded above and below, whatever the sign of $f$. An important question, raised in \cite{Elitzur:2005kz}, is whether quantum corrections stabilize the potential. One can readily check that, unlike in the $D=4, \,{\cal N}=4$ SYM theory studied in \cite{Witten:2001ua} and used for cosmology in \cite{Craps:2007ch}, the beta function for the coupling $f$ vanishes to leading order in the $1/N$ expansion. However, at next to leading order we find corrections from the diagrams in figures~\ref{2loop} and~\ref{4loop}. 

It is important to know whether ABJM theory deformed by the triple trace potential \eq{triple} can be defined without a UV cutoff (and if a UV cutoff is necessary, whether it influences the dynamics of interest). Since a complete analysis appears rather complicated we shall, as in Ref.~\cite{Elitzur:2005kz}, begin by studying simpler but analogous scalar field models sharing key features with the theory of interest. In fact, in the limit of weak 't~Hooft coupling, $N/k\to 0$, our deformation of ABJM theory precisely reduces to the $O(2N^2)\times O(2N^2)$ vector model at large $N$. In section~\ref{sub:vector}, we first discuss the $O(N)$ vector model, drawing an important distinction between the cases with positive and negative coupling. In section~\ref{sub:vector2}, we then study the $O(N)\times O(N)$ vector model, which appears as the weak 't~Hooft coupling limit of the deformed ABJM model of interest. Finally, in section~\ref{sub:comments}, we comment on the case of non-zero 't~Hooft coupling. 

\subsection{The $O(N)$ vector model}\label{sub:vector}
\begin{figure}
\begin{center}
\epsfig{file=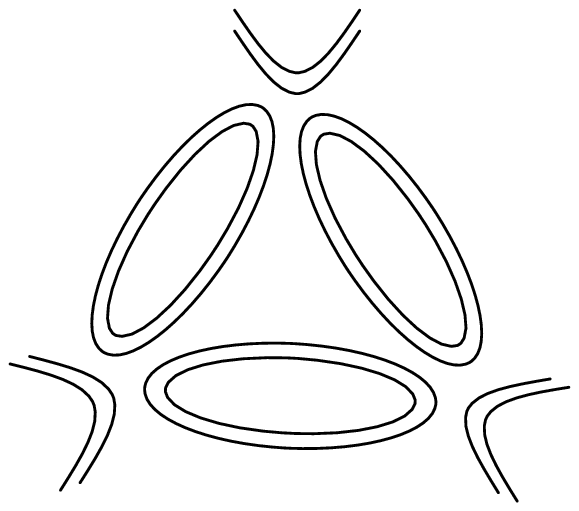, width=8cm}
\end{center}
\caption{Four-loop diagram that renormalizes the coupling $f$ at order $1/N^2$.}
\label{4loop}
\end{figure}

Before discussing the subleading corrections in the model of interest, let us first discuss the analogous question in the well-understood $O(N)$ vector model at the tri-critical point.
The latter model, which describes $N$ scalar fields in three dimensions, assembled in a vector $\vec\phi$, is defined by the action
\be\label{vectormodel}
S=\int d^3x\left(
-\half\partial_\mu\vec\phi\cdot\partial^\mu\vec\phi-{1\over 6} {\lambda\over N^2}\left(\vec\phi\cdot\vec\phi\right)^3 
\right).
\ee
The sextic vertex is the analogue of figure~\ref{vertex6}, with all double lines replaced by single lines. The perturbative beta function for $\lambda$ vanishes to leading order in the $1/N$ expansion, but receives nonzero contributions of order $\lambda^2/N$ and $\lambda^3/N$ from the logarithmically divergent two- and four-loop diagram analogous to figures~\ref{2loop} and~\ref{4loop} (with all double lines replaced by single lines). (Contributions with higher powers of $\lambda$ are suppressed by additional powers of $1/N$.) The sum of the Feynman diagrams in figures~\ref{vertex6}, \ref{2loop} and~\ref{4loop} is \cite{Stephen_McCauley, Lewis_Adams, Pisarski:1982vz}
\be\label{diagrams}
3!\,2^3\,\left[
-i{\lambda\over 6 N^2}+{i\over 2}\ln\left({\Lambda^2\over p^2}\right){\lambda^2\over 36 N^4}{9N\over\pi^2}-{i\over2}\ln\left({\Lambda^2\over p^2}\right){\lambda^3\over 216 N^6}{9N^3\over32\pi^2}
\right],
\ee
times the appropriate tensor, where $\Lambda$ is a UV cutoff and the scale $p^2$ is set by (spacelike) momenta flowing in and out of the diagrams. From \eq{diagrams}, one reads off the beta function
\be\label{beta}
\beta(\lambda)={3\over 2 \pi^2 N}\left(\lambda^2-{\lambda^3\over 192}\right),
\ee
which is indeed suppressed by $1/N$. 

For a positive potential ($\lambda>0$), it follows from the quadratic term in \eq{beta} that the coupling is marginally irrelevant for small values of $\lambda$, for which it increases towards the UV. As $\lambda$ increases, the cubic term in \eq{beta} becomes important and a perturbative UV fixed point is reached at
\be
\lambda^*=192.
\ee 
In \cite{Bardeen:1983rv}, a self-consistent, static, non-perturbative UV fixed point was found, in the strict $N=\infty$ limit, at the smaller value
\be
\lambda_c={16\pi^2 }<\lambda^*,
\ee
and an instability was established for $\lambda>\lambda_c$, meaning that if one attempts to construct a static vacuum, all masses are of the order of the cutoff, so that the theory does not possess a continuum limit. See \cite{Elitzur:2005kz} for a recent discussion. However, preliminary results indicate that there is no such UV-dependence in the time-dependent backgrounds of interest to us~\cite{CHTinprogress}. 

For a negative potential ($\lambda<0$), the quadratic term in \eq{beta} implies that the coupling is asymptotically free (as discussed in \cite{Coleman:1973sx} for $-\phi^4$ theory in four dimensions). As mentioned in \cite{Coleman:1973sx} (see appendix~B of \cite{Craps:2007ch} for a recent discussion in a context closely related to the present paper, and the discussion below), one can then use the techniques of \cite{Coleman:1973jx} to show directly that the energy of the system is unbounded below.\footnote{Based on this the authors of \cite{Coleman:1973sx} dismissed the theory as nonsense, whereas in \cite{Craps:2007ch} a first attempt was made to make sense of such field theories. An update on the latter work will appear elsewhere \cite{CHTinprogress}.} So at least for the $O(N)$ vector model, the fact that the potential with $\lambda<0$ is unbounded below definitely survives quantum corrections.%
\footnote{
This conclusion might at first sight appear different from that reached in \cite{Elitzur:2005kz}. However, looking more closely, the computations referred to in \cite{Elitzur:2005kz} refer to {\em positive} coupling (beyond the critical one). 
}

We now compute the Coleman-Weinberg effective potential in the regime $-1\ll\lambda<0$, so that the first term in \eq{beta} dominates. The coupling as a function of the renormalization scale $\mu$ is determined by the Callan-Symanzik equation
\be
\mu{d\lambda\over d\mu}={3\lambda^2\over 2\pi^2N},
\ee
whose solution is 
\be\label{fmu}
\lambda_\mu=-{4\pi^2N\over 3\ln(\mu^2/M^2)},
\ee
with $M$ being an arbitrary mass scale (implementing dimensional transmutation). The Coleman-Weinberg potential, where the renormalization scale is set by a field value, is then
\be\label{2looppotential}
V(\vec\phi)=-{4\pi^2\over 3N\ln\left[\left(\vec\phi\cdot\vec\phi\right)/M^2\right]}\,\left(\vec\phi\cdot\vec\phi\right)^3.
\ee
If $M$ is chosen such that the coupling $\lambda$ is small, $-1\ll\lambda<0$, for some value of $\vec\phi\cdot\vec\phi$, it will be even smaller for larger values. (Note that the condition $-1\ll\lambda<0$ implies that the logarithm in \eq{2looppotential} should be at least of order $N$.) Therefore, the perturbative analysis leading to the potential \eq{2looppotential} is reliable for sufficiently large field values, which establishes that it is unbounded below. 

\subsection{The $O(N)\times O(N)$ vector model}\label{sub:vector2}

Consider the $O(N)\times O(N)$ vector model defined by the action
\bea\label{vectormodel2}
S&=&\int d^3x\left[
-\half\partial_\mu\vec\phi_1\cdot\partial^\mu\vec\phi_1-\half\partial_\mu\vec\phi_2\cdot\partial^\mu\vec\phi_2-{\lambda_{111}\over 6 N^2}\left(\vec\phi_1\cdot\vec\phi_1\right)^3 - {\lambda_{222}\over 6 N^2}\left(\vec\phi_2\cdot\vec\phi_2\right)^3\right.\cr
&&\left. \ \ \ \ \ \ \ \ \ - {\lambda_{112}\over 6 N^2}\left(\vec\phi_1\cdot\vec\phi_1\right)^2\left(\vec\phi_2\cdot\vec\phi_2\right)- {\lambda_{122}\over 6 N^2}\left(\vec\phi_1\cdot\vec\phi_1\right)\left(\vec\phi_2\cdot\vec\phi_2\right)^2
\right].
\eea
For the special case $\lambda_{112}=-\lambda_{122}=-3\lambda_{111}=3\lambda_{222}\equiv -3\lambda$, this corresponds to the potential
\be\label{ratio}
V={\lambda\over 6 N^2}\left(\vec\phi_1\cdot\vec\phi_1-\vec\phi_2\cdot\vec\phi_2\right)^3.
\ee
By collecting the $2N^2$ real components of the complex $N\times N$ matrix $Y^1$ in a $2N^2$-component vector $\vec\phi_1$, and similarly for $Y^2$, we see that the triple trace potential \eq{triple} takes the form \eq{ratio} (with $N$ replaced by $2N^2$). Moreover, in the $N/k\to 0$ weak 't~Hooft coupling limit, the deformed ABJM action precisely reduces to that of the $O(2N^2)\times O(2N^2)$ vector model.

When we consider the potential \eq{ratio}, we see that there are four terms, and that the potential does not have a definite sign. In fact, even if they appear in fixed ratios in the classical potential \eq{ratio}, the couplings of the four terms will renormalize differently, which is why we include four different couplings in \eq{vectormodel2} to investigate the ultraviolet properties of the theory with potential \eq{ratio}.%
\footnote{\label{mixing}
Incidentally, at large 't~Hooft coupling, one would use the bulk description to investigate the renormalization properties of the triple trace potential \eq{triple}. Since the field $\varphi$ dual to the operator ${\cal O}$ is part of the consistent truncation \eq{trunc}, it does not source the other scalar fields and one might be tempted to conclude that at large 't~Hooft coupling, unlike what happens at weak 't~Hooft coupling, the potential \eq{triple} preserves its form (\ref{ratio}) under renormalization group flow. However, as in \cite{Gubser:2002zh}, pure AdS continues to be a solution to the classical equations of motion of \eq{trunc} even with modified boundary conditions \eq{bc}, which indicates that there is no running at the level of classical supergravity. (Moreover, unlike the double trace deformation analyzed in \cite{Gubser:2002zh}, the modified boundary conditions preserve the asymptotic AdS symmetry group.) It therefore seems to us that seeing any non-trivial renormalization group flow will require going beyond the classical supergravity approximation, and for this it is important to know whether properties related to consistent truncation survive quantum corrections in the bulk.
}

It is straightforward to generalize the perturbative beta function \eq{beta} at order $1/N$ to the model \eq{vectormodel2}. The result is
\bea\label{beta2}
&&{2 \pi^2 N \over 3}\beta_{111}=\lambda_{111}^2+{1\over9}\lambda_{112}^2-{1\over 192}\left(\lambda_{111}^3+{1\over3}\lambda_{111}\lambda_{112}^2+{1\over9}\lambda_{112}^2\lambda_{122}+{1\over27}\lambda_{122}^3\right);\cr
&&{2 \pi^2 N \over 3}\beta_{112}={1\over9}\lambda_{112}^2+{1\over9}\lambda_{122}^2+{2\over3}\lambda_{111}\lambda_{112}+{2\over9}\lambda_{112}\lambda_{122}\cr
&& \quad \qquad -{1\over 192}\left(\lambda_{111}^2\lambda_{112}+{2\over3}\lambda_{111}\lambda_{112}\lambda_{122}+{1\over9}\lambda_{112}^3+{1\over3}\lambda_{112}^2\lambda_{222}+{2\over9}\lambda_{112}\lambda_{122}^2+{1\over3}\lambda_{122}^2\lambda_{222}\right);\cr
&&{2 \pi^2 N \over 3}\beta_{122}={1\over9}\lambda_{122}^2+{1\over9}\lambda_{112}^2+{2\over3}\lambda_{222}\lambda_{122}+{2\over9}\lambda_{122}\lambda_{112}\cr
&&\quad\qquad-{1\over 192}\left(\lambda_{222}^2\lambda_{122}+{2\over3}\lambda_{222}\lambda_{122}\lambda_{112}+{1\over9}\lambda_{122}^3+{1\over3}\lambda_{122}^2\lambda_{111}+{2\over9}\lambda_{122}\lambda_{112}^2+{1\over3}\lambda_{112}^2\lambda_{111}\right);\cr
&&{2 \pi^2 N \over 3}\beta_{222}=\lambda_{222}^2+{1\over9}\lambda_{122}^2-{1\over 192}\left(\lambda_{222}^3+{1\over3}\lambda_{222}\lambda_{122}^2+{1\over9}\lambda_{122}^2\lambda_{112}+{1\over27}\lambda_{112}^3\right).
\eea

\begin{figure}
\begin{center}
\epsfig{file=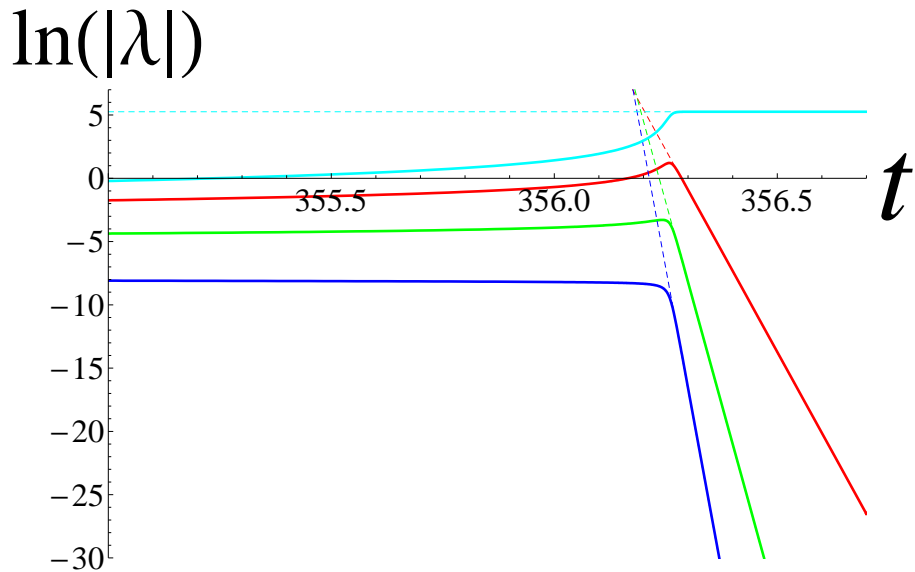, width=12cm}
\end{center}
\caption{Numerical solution for the running couplings in the $O(N)\times O(N)$ vector model, shown against $t\equiv 3\ln (\mu)/(2 \pi^2 N) $. Initial conditions were specified by (\ref{ratio}), with $\lambda=-1/500$, at $t=0$. From top to bottom, the solid curves show $\lambda_{222}$, $\lambda_{122}$, $\lambda_{112}$ and $\lambda_{111}$. The dashed lines show the limiting behavior of the couplings in the UV, given in (\ref{elim}). Note that the absolute values of the couplings are shown: 
$\lambda_{111}$ and $\lambda_{122}$ are negative whereas $\lambda_{112}$ and 
$\lambda_{222}$ are positive.}
\label{runningfig}
\end{figure}

From our knowledge of the $O(N)$ vector model, reviewed above, we can immediately infer the existence of the following perturbative fixed points. One fixed point simply corresponds to the non-trivial UV fixed point of the $O(2N)$ vector model:  $\lambda_{112}=\lambda_{122}=3\lambda_{111}=3\lambda_{222}=3\lambda^*$. A second fixed point has all couplings equal to zero; it can be approached in the UV by starting with the $O(2N)$ vector model with small negative coupling. The perturbative fixed point of interest to us has $\lambda_{222}=\lambda^*$ and $\lambda_{112}=\lambda_{122}=\lambda_{111}=0$. By integrating \eq{beta2} numerically (see figure~\ref{runningfig}), we find that this UV fixed point is reached when we start with \eq{ratio} and flow towards the UV. If we choose conventions such that $\lambda<0$ in \eq{ratio} (the other sign is related to this by interchanging $\vec\phi_1$ and $\vec\phi_2$), $\lambda_{111}$ remains negative and approaches zero as the renormalization scale is increased, just as we encountered in the $O(N)$ vector model with negative coupling. From (\ref{beta2}), one determines the detailed UV-limiting behavior of the four couplings,
\bea
\label{elim}
\lambda_{111} \rightarrow {1\over 27 (192)^2} \lambda_{122}^3, \ \lambda_{112}\rightarrow {1\over 576} \lambda_{122}^2, \ \lambda_{122} \rightarrow C \mu^{-(96/\pi^2 N)},\
\lambda_{222} \rightarrow 192, \ \mu \rightarrow \infty,
\eea
with $C$ a negative constant (see Fig. \ref{runningfig}). This behavior will be important in our discussion of the related cosmology\cite{CHTinprogress}.

As we have already mentioned, it is known that, at large $N$, non-perturbative effects destabilize the model for positive coupling $\lambda > \lambda_c$, so that it does not possess a UV-independent, static ground state~\cite{Bardeen:1983rv,Rabinovici:1987tf}.
However, for describing cosmology, and cosmological singularities in particular, we are not interested in static ground states, and whether any UV-dependence enters depends on the questions being asked. We shall detail this point in future work~\cite{CHTinprogress}, where we use the model presented here to study
the cosmological space-times mentioned in the introduction.

\subsection{Triple trace deformation of ABJM theory: comments}\label{sub:comments}

In the previous subsection, we have studied the vector model arising in the limit of zero 't~Hooft coupling, in which we could ignore the single trace interactions in the ABJM action \eq{ABJMaction}.  As in \cite{Craps:2007ch}, the bulk is in a stringy regime for weak 't~Hooft coupling. However, from the bulk analysis described in section~\ref{HH}, which is valid at large 't~Hooft coupling, we know that at least certain important features, such as the unboundedness of the potential and the absence of logarithmic running to leading order in $1/N$, extend to the regime with large 't~Hooft coupling. Another question one may ask is whether the beta functions of the deformation couplings will receive corrections linear in $f$, for instance proportional to $f/k^2$ rather than $f^2$. However, such logarithmically divergent diagrams need to cancel because ${\cal O}$ is a chiral primary operator, whose anomalous dimension must vanish in the undeformed theory: from any logarithmically divergent diagram with one triple trace and one single trace vertex contributing to the triple trace coupling, one can construct a logarithmically divergent diagram contributing to the anomalous dimension of ${\cal O}$ by stripping off two uncontracted $\Tr(YY^\dagger)$ factors from the triple trace vertex. 

An important difference with the $D=4,\, {\cal N}=4$ SYM theory, where the running of $f$ occurred at leading (zeroth) order in the $1/N$ expansion, is that here the running of $f$ is suppressed by $1/N^2$. This is consistent with the absence of logarithmic terms in the asymptotic bulk supergravity solutions, mentioned at the end of section~\ref{HH}, and can therefore be regarded as a test of the AdS/CFT correspondence.

\section*{Acknowledgments}
We would like to thank A.~Bernamonti, J.~de Boer, I.~Klebanov and D.~Tong for useful discussions, and E.~Rabinovici for comments on the manuscript. In addition, we thank the anonymous PRD referee for comments leading to footnote~\ref{mixing} and to an improved discussion in section~4. B.C.\ and T.H.\ are grateful for the hospitality of the Centre for Theoretical Cosmology in Cambridge and Perimeter Institute for Theoretical Physics at several stages of this project. B.C.\ also thanks the Galileo Galilei Institute for Theoretical Physics for hospitality and INFN for partial support during the completion of this work. The work of B.C.\ was supported in part by the Belgian Federal Science Policy Office through the Interuniversity Attraction Pole IAP VI/11 and by FWO-Vlaanderen through project G.0428.06.  The work of N.T.\ is supported by the Perimeter Institute for Theoretical Physics.

\end{document}